# Multi-scale plasticity homogenization of Sn-3Ag-0.5Cu: from β-Sn micropillars to polycrystals with intermetallics

Yilun Xu[1]*, Tianhong Gu[1,2,]*, Jingwei Xian[1], Finn Giuliani[1], T. Ben Britton[1.3], Christopher M. Gourlay[1], and Fionn P.E. Dunne[1]

[1]Department of Materials, Imperial College London, SW7 2AZ, UK.

[2]School of Metallurgy and Materials, University of Birmingham, B15 2TT, UK.

[3]Department of Materials Engineering, University of British Columbia, Vancouver, British Columbia, V6T 1Z4

**Abstract**

The mechanical properties of β-Sn single crystals have been systematically investigated using a combined methodology of micropillar tests and rate-dependent crystal plasticity modelling. The slip strength and rate sensitivity of several key slip systems within β-Sn single crystals have been determined. Consistency between the numerically predicted and experimentally observed slip traces has been shown for pillars oriented to activate single and double slip. Subsequently, the temperature-dependent, intermetallic-size-governing behaviour of a polycrystal β-Sn-rich alloy SAC305 (96.5Sn-3Ag-0.5Cu wt%) is predicted through a multi-scale homogenization approach, and the predicted temperature- and rate-sensitivity reproduce independent experimental results. The integrated experimental and numerical approaches provide mechanistic understanding and fundamental material properties of microstructure-sensitive behaviour of electronic solders subject to thermomechanical loading, including thermal fatigue.

---

* Corresponding author: t.gu@bham.ac.uk; yilun.xu@imperial.ac.uk

# 1 Introduction

This paper proposes an anisotropic, intermetallic size-dependent, multi-scale crystal plasticity homogenization model that predicts the mechanical behaviours of a β-Sn-rich solder alloy SAC305 (96.5Sn-3Ag-0.5Cu wt%). The rate-dependent crystal plasticity of β-Sn, contributing as the major phase of the SAC305 alloy, has been established through systematic single crystal micropillar testing, characterisation and crystal plasticity modelling with a range of crystallographic orientations under various loading rates at room temperature. Slip system properties have been extracted based on the experimental and crystal plasticity simulated load-displacement response, and both single and double slip have been captured such that the plastic anisotropy is investigated in depth. A polycrystal SAC305 representative volume element that employs the material properties established from the multi-scale homogenization captures the rate-sensitive, temperature-dependent mechanical behaviour, which is compared to independent macro-scale experiments. The research provides the link from single crystal properties through to thermomechanical response of solder components.

The family of Sn-Ag-Cu (SAC) alloys are currently widely used as Pb-free solders [1]. An important and extensively used member of the SAC solder family is SAC305 (96.5Sn–3.0Ag-0.5Cu, wt%), whose performance has been shown to be strongly correlated to its microstructure [2, 3] when subject to in-service loading conditions, *e.g.* thermal cycling and shock. β-Sn is the major phase within the SAC305 alloy and detailed understanding of the mechanical properties (including the moduli, slip strengths, hardening and rate sensitivity etc) of β-Sn crystals remains important, along with mechanistic understanding of the failure mechanisms at the microstructural scale for solders. Mechanical tests [4, 5] and modelling efforts [6] on polycrystal SAC305 specimens have been conducted, such that isotropic, macroscopic material properties have been obtained and used to quantify behaviour in an averaged, homogenised sense. However, the strong anisotropic behaviour of solder joints [7-11] that commonly contain only one to three β-Sn grains [12-16] requires the determination of the anisotropic single crystal properties for β-Sn and SAC305.

Micropillar compression tests have been extensively used to understand and extract the micromechanical properties for various single crystal materials [17], such as those for nickel [18], copper [19] and titanium alloys [20]. Earlier experimental efforts [21-23] have investigated micro-deformation in β-Sn single crystals in micropillar compression, but investigations have not yet delivered consistent measurements of the critical resolved shear stresses (CRSSs) for individual slip system. Consistency is difficult to achieve, as a variety of model have been employed, and in addition, the strain rate sensitivity of β-Sn single crystals which dominates the creep behaviour of solders under thermal cycling [8, 24], remains to be fully quantified at the crystallographic slip scale. Together, these prevent development of anisotropic rate-dependent crystal plasticity modelling of β-Sn and SAC305 alloys which is required to engineer reliable models to support the confident and cost-effective use of these alloys in safety critical systems.

This paper begins reporting of micropillar compression experiments where the load-displacement response, crystal orientation, and slip traces are captured from experimental tests of single crystal β-Sn pillars. These results are then used within a crystal plasticity model to enable the material properties to be extracted. A multi-scale modelling methodology which recognises the key features of SAC305 alloy microstructures, including IMC-containing eutectic regions and β-Sn dendrites, is developed through homogenisation such that the mechanical behaviour of Sn-based solders in service [25] may be captured. The homogenised model is quantitatively assessed by comparing temperature- and strain rate- sensitivity of a multi-scale homogenised SAC305 polycrystal representative volume element (RVE) with independent tension tests.

The research sheds new light on the following aspects:

- The anisotropic, rate-dependent β-Sn material properties have been determined through combining the single crystal micropillar experiments and CPFE modelling. The literature data on the β-Sn properties has shown strong scattering and no direct crystal plasticity model has yet been conducted.

- The anisotropic material properties of the SAC305 alloy have then been determined through a multi-scale modelling strategy, with the knowledge from the micropillar tests and considering the detailed microstructure, such as intermetallic-containing eutectic regions.

- The obtained SAC305 properties have been validated through a systematic comparison to several independent tensile tests under various loading rates and temperatures. Therefore, it provides the fundaments to investigate the thermo-mechanical behaviours of SAC305 solder joints subject to in-service conditions.

## 2 Experimental investigation

### 2.1 Sample preparation and loading

Micropillars were fabricated within bulk 99.99% tin (HP Sn) samples, which is β-Sn at room temperature. Two samples were prepared; sample 1 in Figure 1(a) is a poly-crystal sample aged at 100°C for 480 h with several large grains, and sample 2 in Figure 1(b) is a bi-crystal sample produced by directional solidification containing two twin grains (the red grain is the mechanical twin grain of the cyan grain formed by shear cutting across the sample, Figure 1(b)). To achieve activation of single slip for a particular slip system, orientations indicated by yellow and pink (Figure 1(a)) were selected from sample 1. There is a significant difference in Schmid factor values between the first and the second activated slip systems for these two orientations (see Appendix Figure A 1 for the Schmid factor analysis for the two grains). For sample 2, the cyan grain has near-[100] orientated in the loading direction (Figure 1 (b)), and a 23° angular deviation of its [100] to the loading direction. The red grain has a near-[001] in the loading direction (Figure 1(b)) with a 22° angular deviation of its [001] to the loading direction. The crystallographic orientations of the samples were determined using electron backscatter diffraction (EBSD) with a step size of 6 μm. The sample surface was prepared by mechanical polishing with a 0.05 μm colloidal silica and followed by a refinement using a broad Ar ion beam milling (Gatan PECS II) at 0.5 keV with an incident angle of 1° for 10 minutes.

An array of square-section micropillars of ~5 μm mid-height width and ~13 μm height (Figure 1 (c)) with dimensions and taper angles given in Table 1 were cut using Thermo Scientific NanoBuilder scripting within the selected four grains by Ga$^+$ FIB milling at an accelerating voltage of 30 kV and reducing ion beam current from 9 nA to 280 pA using a FEI Helios Nanolab 600 Dual beam microscope to achieve an aspect ratio (*i.e.* width/length) of 1:2.5. The low current beam was used for the final milling of the micropillar contour to minimise the damage of the surface orientation. The dimensions and crystallographic orientations of the single crystal pillars are listed in Table 1. The sizes of the fabricated pillars were chosen to negate the size effect that was observed in other micromechanical tests [23, 26].

The *in situ* uniaxial micropillar compression tests were set up within a Quanta FEG 450 SEM using a displacement-controlled Alemnis nanoindentation stage with a 10 μm diameter circular flat punch indenter. The pillars were deformed at room temperature with the applied displacements, loading and unloading (displacement) velocities, engineering strains and loading rate given in Table 2. Further experimental details can be found in [27].

Table 1. Dimensions and crystallographic orientations of the four investigated pillars

|  |  | Pillar 1 | Pillar 2 | Pillar 3 | Pillar 4 |
|---|---|---|---|---|---|
| $a_1 (\mu m)$ | | 4.50 | 4.61 | 4.60 | 4.75 |
| $a_2 (\mu m)$ | | 5.83 | 5.57 | 5.50 | 5.67 |
| $h (\mu m)$ | | 11.92 | 13.77 | 13.01 | 13.31 |
| $\omega (°)$ | | 3.17 | 1.99 | 1.98 | 1.98 |
| Euler angles (°) | $\psi 1$ | 2.4 | 160.4 | 107.4 | 152.4 |
|  | $\theta$ | 38.5 | 35.5 | 77.8 | 22.0 |

|  | ψ2 | 35.6 | 218.5 | 250.4 | 217.7 |

Table 2. Summary of the mechanical data of the four investigated pillars. The average engineering strain ($\varepsilon_{eng}$) was calculated from displacement/height of the micropillar.

|  | Pillar 1 | | Pillar 2 | | Pillar 3 | | Pillar 4 | |
| --- | --- | --- | --- | --- | --- | --- | --- | --- |
| Displacement (μm) | 0.6 | 1.2 | 0.6 | 1.2 | 0.65 | 1.3 | 0.65 | 1.3 |
| Max. engineering strain (%) | 5 | 10 | 5 | 10 | 5 | 10 | 5 | 10 |
| Loading velocity (μm s$^{-1}$) | 0.005 | 0.01 | 0.005 | 0.01 | 0.005 | 0.01 | 0.005 | 0.01 |
| Unloading velocity (μm s$^{-1}$) | 0.05 | 0.1 | 0.05 | 0.1 | 0.05 | 0.1 | 0.05 | 0.1 |
| Loading rate (s$^{-1}$) | $4 \times 10^{-4}$ | $8 \times 10^{-4}$ | $4 \times 10^{-4}$ | $8 \times 10^{-4}$ | $4 \times 10^{-4}$ | $8 \times 10^{-4}$ | $4 \times 10^{-4}$ | $8 \times 10^{-4}$ |

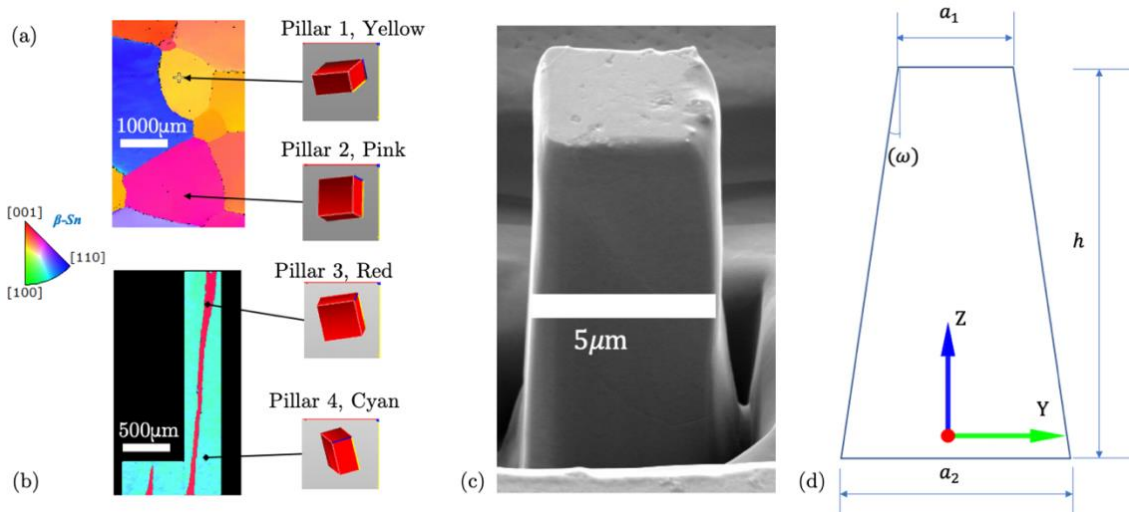

Figure 1. Inverse pole figure (IPF) maps with respect to the compression direction, showing crystallographic orientations of regions chosen for pillar fabrication for (a) Pillars 1 and 2 within sample 1, (b) Pillars 3 and 4 within sample 2. (c) A representative experimental micropillar (Pillar 3 for example) before compression tests. (d) Schematic cross-section of a pillar with key dimensions.

## 2.2 Experimental observation of the micropillars

Figure 2 shows the micrographs of the four pillars after compression, slip bands can be observed on all four faces for each between the top corner and the edge of the micropillar, (Figure A 2) with annotations of the observed experimental slip systems numbered according to the conventions of the crystal plasticity simulations. For pillars 1 and 2, the activation of differing single slip systems is observed on side surfaces indicated with white dashed lines (Figure 2a, b). However, two slip traces are observed on the surface of pillars 3 and 4 after compression (Figure 2c, d), implying activation of two different slip systems. By analysing the experimental slip traces and calculating the values of (uniaxial) Schmid factor, $m=\cos\varphi \times \cos\lambda$, for all slip systems of each pillar, the Schmid factors for slip systems SS-2 and SS-3 (see the definition of slip systems in Table A 1) were found to be the highest of those for pillars 1 and 2, and their experimentally observed slip traces align well with those expected for slip system 2 ((010) [001]) and 3 ((110) [001]) on the pillar side surfaces shown (Figure 2a, b). For pillar 3, slip systems 8 $((110)[\bar{1}11]/2)$ and 10 ($(1\bar{1}0)[\bar{1}\bar{1}1]/2$) are activated since the experimental traces show good alignment with the calculated traces, which have the two highest Schmid factors (Figure 2c). In addition, activation of slip systems 2 and 3 are also expected for pillar 4 from considerations of the Schmid factor ranking for the observed slip traces in Figure 2d [27]. The activated slip system(s) are justified using the crystal plasticity model later in Section 3.3.

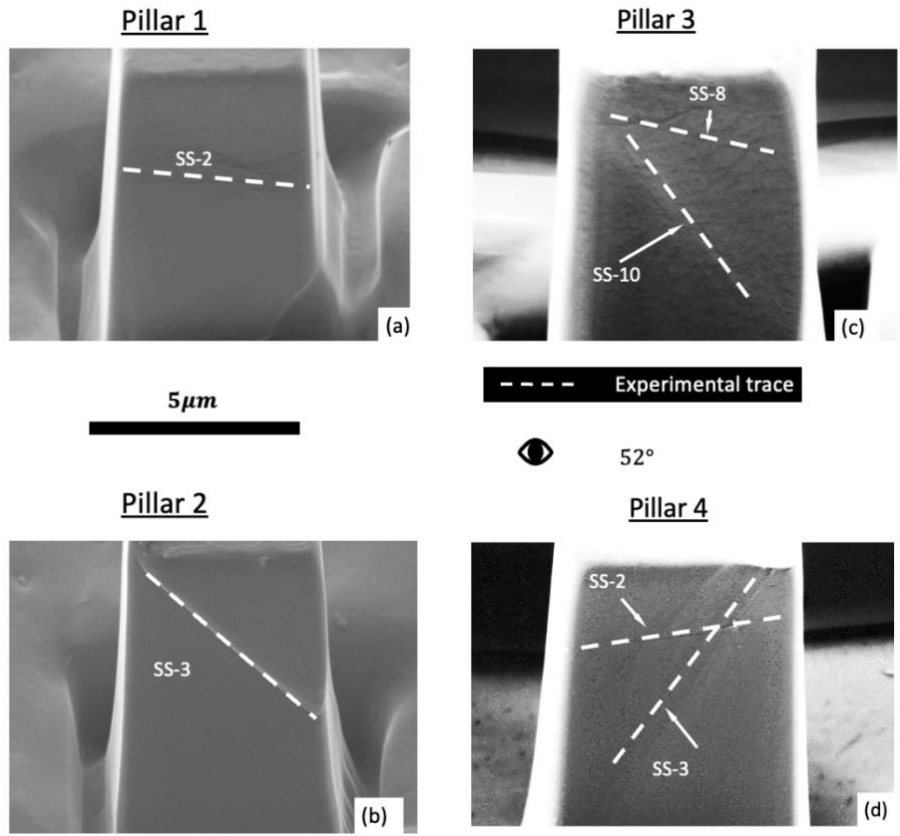

Figure 2. SEM micrographs of deformed micropillars at 5 % strain for (a) pillar 1, (b) pillar 2, (c) pillar 3, (d) pillar 4. One (single) slip trace (a, b) and two (double) slip traces (c, d) are observed experimentally and highlighted using white dashed lines. The activated slip systems are anticipated from Schmid factor calculations.

## 2.3 Mechanical response of the micropillars

Typical force-displacement responses of pillars with different crystallographic orientations are shown in Figure 3 (for pillars 3 and 4 as an example) for the displacement rate of 0.01 μm s$^{-1}$, corresponding to a target nominal strain rate of $\dot{\epsilon}$ =8 × 10$^{-4}$ s$^{-1}$. The reaction forces were obtained by averaging over five adjacent load data points to exclude the noise in the system. Apparent hardening behaviours are observed in both pillars, and the intermittent response of the reaction forces reflect dynamic dislocation and strain bursts [28-30] taking place within the pillars. A strong crystallographic orientation effect is observed, as pillar 3 shows considerably higher force required compared to that for pillar 4, influenced in addition by the anisotropy of slip strengths in β-Sn single crystals [31]. In addition to the plastic anisotropy, the (elastic) unloading curve reveals dependence upon the crystallographic orientation. The elastic constants are extracted from the unloading stage of the force-displacement curve, to reduce the impact of microplasticity during loading and account for compliance reductions due to other items within the loading train (adhesive, sample mounting and load frame) [32]. The material properties of the β-Sn single crystal are extracted from these data by calibration with crystal plasticity modelling.

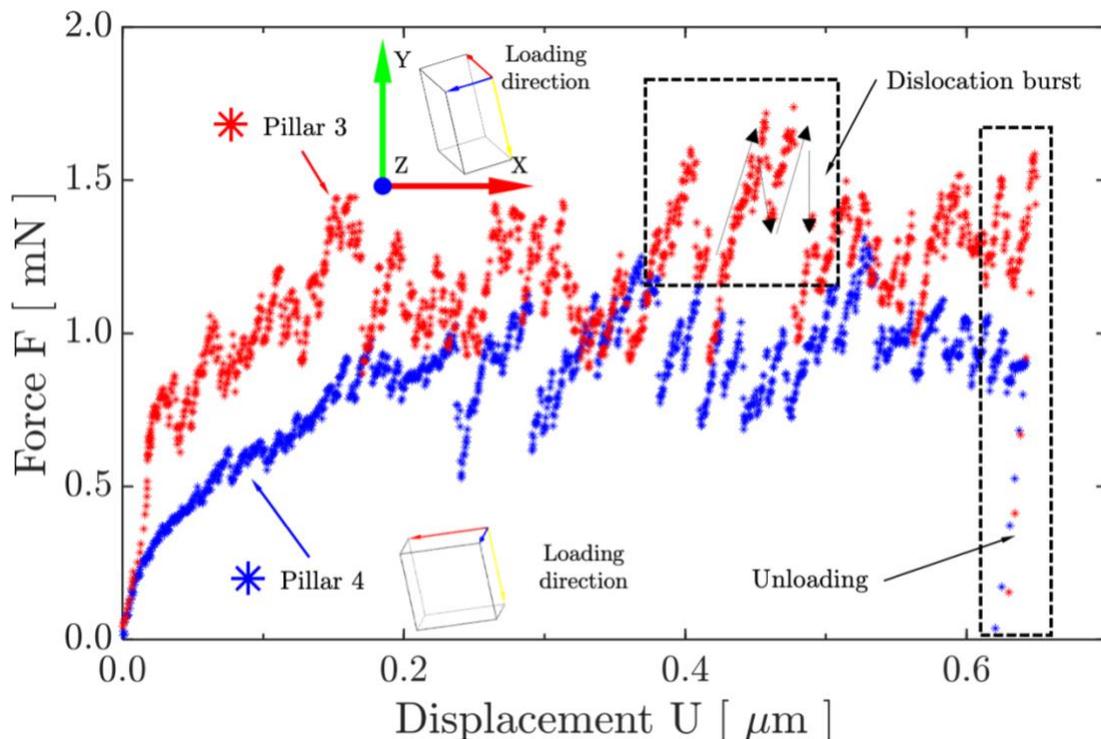

Figure 3. Force-displacement response during loading and unloading obtained from the compression tests with a nominal strain rate $\dot{\epsilon} = 8\times10^{-2}\,s^{-1}$ for pillar 3 and 4 as an example. Loading is in the Z-direction. Unit cells showing the crystallographic orientation of the pillars as indicated in the inset figures.

# 3 Crystal plasticity modelling of the pillars

In this section, the crystal plasticity finite element (CPFE) formulation is briefly introduced for the β-Sn single crystal, and the development of model micropillar compression tests presented. These models are used with the experimental data to determine the β-Sn material properties after which an assessment is presented of the observed slip system activations through comparisons of the experimentally observed and model predicted slip traces.

A rate-dependent, dislocation-based crystal plasticity framework is used for β-Sn [33], and the full details of its formulations are given in Appendix C. The framework reflects the Body Centre Tetragonal (BCT) slip system intrinsic strengths, and the slip system hardening through both SSD and GND evolution (where the latter is determined from plastic strain gradients and the Nye tensor [34]). The strain rate sensitivity for each individual slip system is incorporated through consideration of thermally-activated pinned dislocation escape from obstacles and glide, which is quantified through an activation energy $\Delta F$ associated with an appropriate volume $\Delta V$ related to dislocation escape [35]. The CPFE formulation includes the incorporation of thermal strains (including the anisotropy of thermal expansivity which occurs in β-Sn crystals) as well as the temperature-sensitivity of appropriate material properties, specified later. Full details may be found in [33].

## 3.1 Model set up for the pillar tests

A multi-part finite element model, shown in Figure 4(a), that explicitly represents the contribution of the base and glue to the system compliance is developed to simulate the micropillar tests. The β-Sn single crystal pillar is assigned the corresponding dimensions and crystallographic orientations that were measured in the tests (see Table 1). The dimension of the base is set as sufficiently large (at least 10 times that of the dimensions of the pillar) to be representative of the test, and the base shares the same material properties as the pillar. The glue is taken to be an elastic isotropic solid, whose elastic modulus and Poisson's ratio have been obtained from [32]. It has been found necessary to explicitly model the compliance of the base and glue arising from their influence on the elastic response of the system. The elastic constants of the β-Sn single crystal are determined from the unloading stage as opposed to the loading of the load-displacement response curves. The pillar is discretized by 3,500 C3D20R (2$^{nd}$ order hexagonal) elements with a uniform element size of 0.5μm while the biased mesh becomes coarser (ranging from 0.5 μm to 10 μm) towards the bottom of the base and glue to save computational resource. A mesh sensitivity study was performed to ensure the mesh size adopted is suitable for representing the experiments (see Figure A 3). Displacement boundary conditions reflecting the compression of the micropillar are imposed on the top surface (*i.e.* parallel to the Z-axis as sketched in Figure 4(b)), while the four side surfaces remain traction-free. The bottom face of the glue is constrained along the Z-direction *i.e.,* $U_z$=0, and the centre point only is constrained in the lateral directions additionally *i.e.* $U_x$=$U_y$=0. The total Z-direction reaction force of the system is taken to be the sum of the nodal reaction forces along the top face of the pillar.

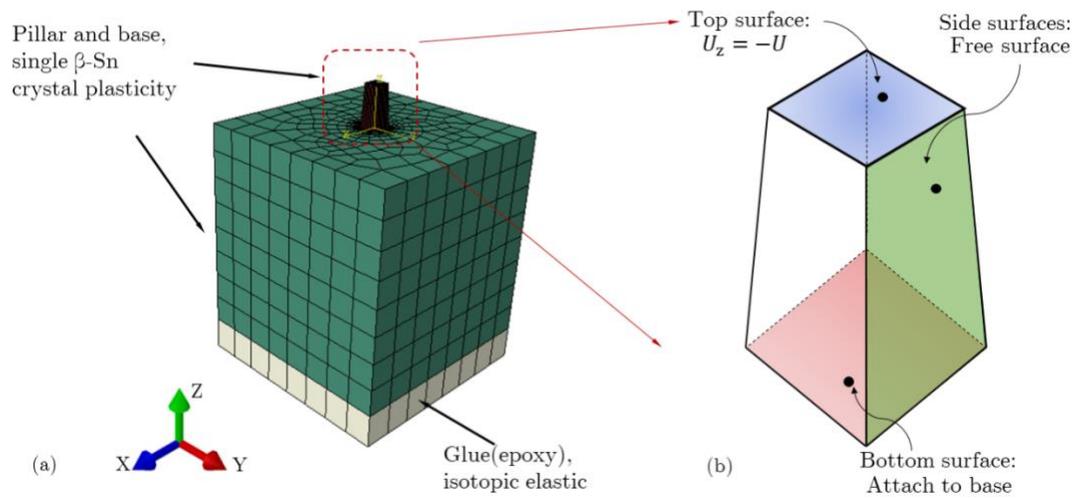

Figure 4. (a) the multi-part crystal plasticity finite element model for simulating the micropillar compression test (b) the schematic boundary conditions imposed on the pillar. The view angle in (b) is slightly rotated to have a better view illustration.

## 3.2 Extraction of slip properties

The crystal slip properties are extracted by matching the modelled micropillar load-displacement curves with those measured from the micropillar tests by virtue of minimizing the errors in the reaction forces simultaneously at both loading rates, which provides the best-fit reflection of the micropillar response in the experiments. The rate sensitivity governing parameters (*i.e.* the activation energy $\Delta H$ and volume $\Delta V$) are assumed to be the same as those that were calibrated from creep tests [33]. Nominal stress-strain response (as opposed to the load-displacement), is calculated where the nominal stress is given by the total Z-direction reaction force divided by the mid-section undeformed area, and the nominal strain is calculated as the Z-displacement of the pillar top face divided by the undeformed height of the pillar. Nominal stress and strain are reported recognising that both the strain and stresses in the pillars are inhomogeneous during compression, because of both the changing pillar section and the strong plastic anisotropy of the β-Sn single crystal [36] and the heterogeneous slip. Figure 5(a) and (b) show a comparison of the 'nominal' stress-strain response as calculated from the experiment and simulation, using the load and displacement data an initial pillar sizes from, for pillars 1 and 2 subject to different loading strain rates, respectively. In addition to the strong anisotropic response in terms of crystallographic orientation (see Figure 3), β-Sn single crystals also exhibit strong rate-sensitivity. By choosing an appropriate set of slip strengths, hardening and rate-sensitivity governing parameters, the CPFE calculated stress-strain response (indicated as lines) captures the experimental results (indicated as scatter data) for the single slip activation cases. It is noted that there are errors at certain strain levels that result from the dislocation dynamic burst [30] events during compression, but the modelled behaviour reflects the overall stress-strain behaviour of the β-Sn single crystals. The resulting properties are listed in Table 3.

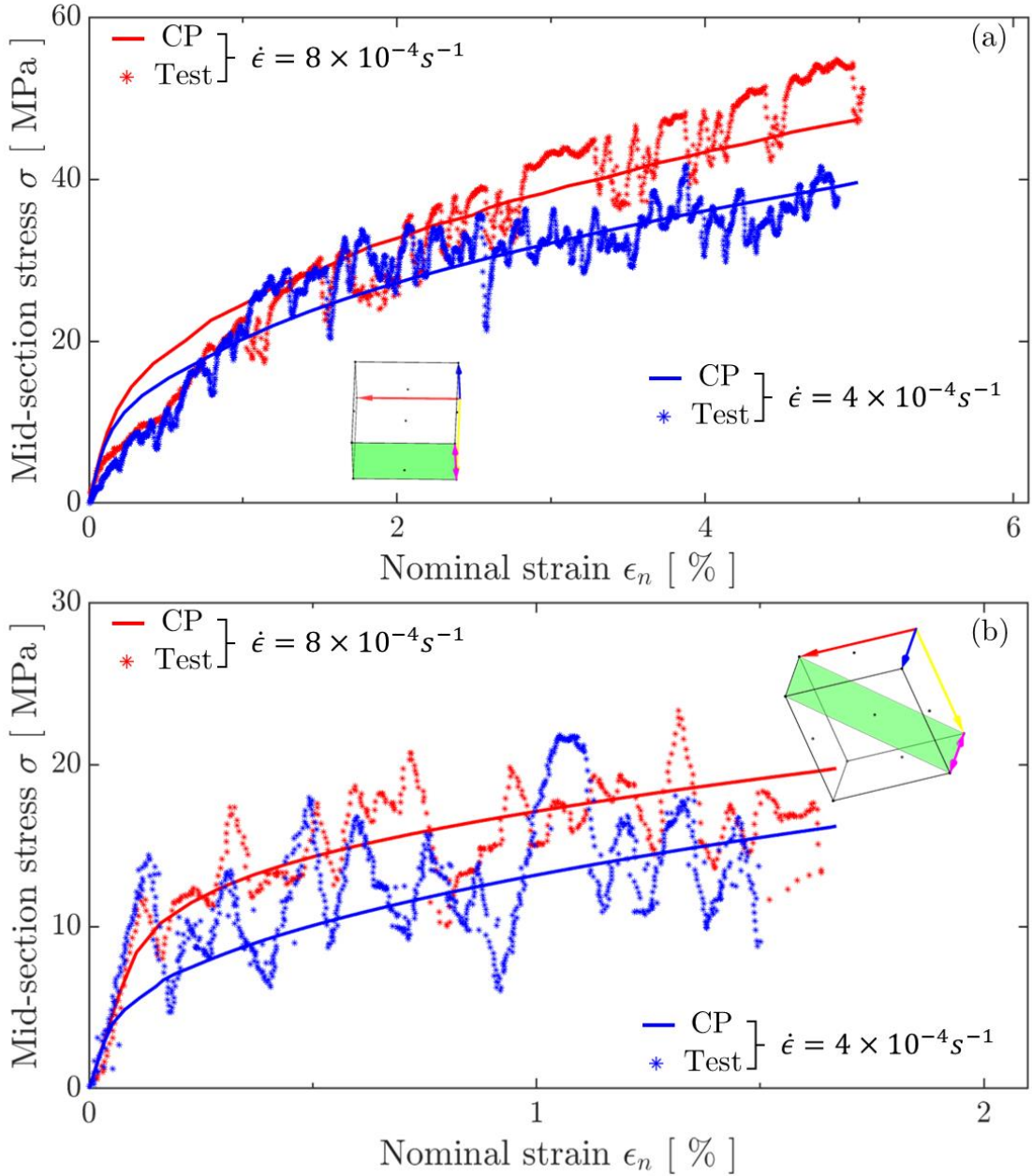

Figure 5. The nominal stress-strain response of (a) pillar 1 and (b) pillar 2 (single slip observed for both) subject to the two loading rates, experimentally measured and calculated using CPFE modelling. The green plane in the unit cell represents the predicted activated slip plane.

The material properties obtained from pillars 1 and 2 by calibrating the elastic constants, slip strengths, hardening coefficients and rate sensitivity parameters for slip systems 2 and 3 are used to model pillars 3 and 4, but the undetermined slip strengths of the other slip systems remain to be calibrated. The stress strain-responses for the two loading rates applied for pillars 3 and 4, are shown in Figure 6 (a) and (b) respectively.

Using the experimental data for Pillar 3, the slip strengths of slip systems 8 and 10 are extracted, and the rate-sensitivity governing parameters $\Delta H$ and $\Delta V$ are found to be identical as for slip systems 2 and 3. For the case of pillar 4, the double slip comes from slip systems 2 and 3, which have been calibrated from the single crystal pillars 1 and 2 tests. The reasonable consistency between the CPFE calculated and experimentally observed stress-strain behaviour further validates the material properties. There is no apparent additional hardening arising from the double slip. Hence, this suggests that double slip in β-Sn single crystals does not evoke dislocation activities associated with cross (latent) hardening.

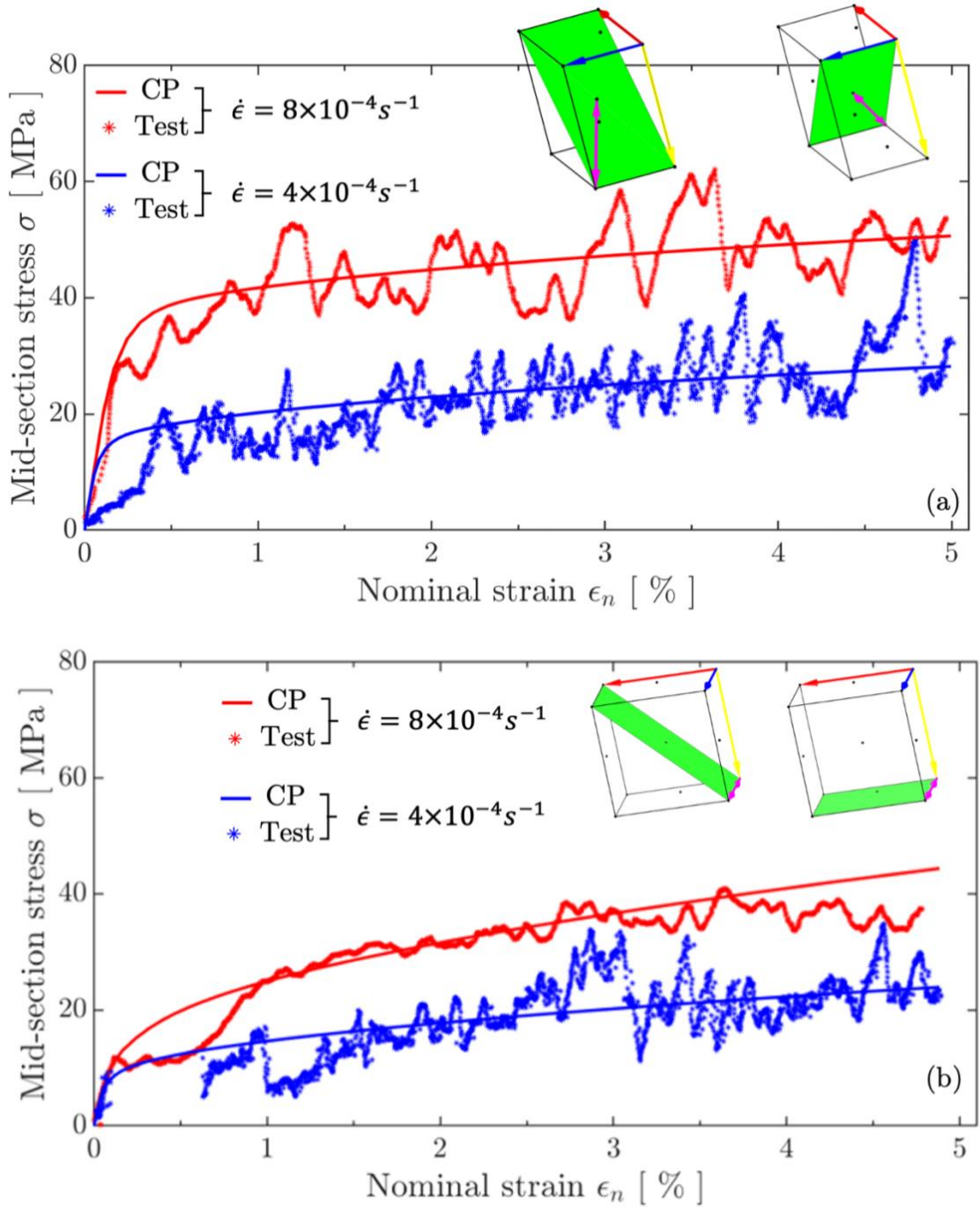

Figure 6. Nominal stress-strain response of (a) pillar 3 and (b) pillar 4 subject to the two loading rates, showing experimentally measured and CPFE computed results.

The crystal plasticity material properties for β-Sn crystals extracted by calibration to the load-displacement response of the pillars are summarized in Table 3; these properties are utilized later in models which reflect IMC content to predict the latter's size and morphology effects on the creep response of directional-solidified SAC305 samples. The calibrated elastic constants are obtained from the unloading load-displacement data, which excludes the contribution from the environment [32], and are found to be close to the values used in previous studies [37] on solder performance.

Bieler and co-workers have previously carried out slip analysis of SAC305 alloys, which contains multi-phases *i.e.* β-Sn and intermetallics ($Ag_3Sn$ and $Cu_6Sn_5$), but they did not explicitly modell the properties of pure β-Sn [7]. Here, we address the more fundamental material properties and mechanical response of the pure β-Sn phase. The plastic anisotropy in β-Sn that manifests itself as the differences between slip system strengths has been found to be different than that of the homogenised SAC305 model used in the literature [38, 39]. The current results reflect the quite large range in property values which have been reported in other studies [21, 22, 40-42], and further highlights the importance of crystallographic orientation of the β-Sn phase on the performance of SAC305 solders subject to in-service loading. However, the strong rate-sensitivity of β-Sn single crystals has been captured, and this is important because other studies [43] argue that the time-dependent behaviour is key to controlling cyclic damage of solder joints subject to in-service conditions.

Table 3. The crystal plasticity material properties for β-Sn crystals extracted from the pillar tests.

| $E_{11}$(GPa) | $E_{33}$(GPa) | $G_{12}$(GPa) | $G_{13}$(GPa) | $\tau_{c0}^{(2,3)}$(MPa) | $\tau_{c0}^{(8,10)}$(MPa) |
|---|---|---|---|---|---|
| 25 | 69 | 22 | 22 | 2 | 6 |
| $\lambda_1(\mu m^{-2})$ | $\lambda_2(\mu m^{-2}s^{-1})$ | $\rho_{ssdm}(\mu m^{-2})$ | $k$(JK$^{-1}$) | $\Delta H$ | $\Delta V$ |
| 200 | 5 | 1.2 | $1.38 \times 10^{-23}$ | 41kJ/0.42eV | $9.26b^3$ |

## 3.3 Identification of activated slip systems

There are three different ways to identify potential slip activation on the free surfaces of the micro-pillars based on slip traces:

(i) the direct experimental observation of slip traces against known crystal orientation through SEM imaging (see Figure 2), which are revealed due to localised out of plane displacement components of the accumulated slip on the imaged surface.

(ii) the analytical calculation from consideration of Schmid factor assuming uniaxial loading. The analytical traces are determined by identifying those systems which have been activated and extracting their normal vectors to calculate the line of intersection generated with the sample free surfaces. The line of intersection is given by $l = n_{sp} \times n_{sf}$ where $n_{sp}$ and $n_{sf}$ are the normal vectors of the slip plane and sample free surface considered, respectively.

(iii) the numerical calculation from CPFE prediction. The crystal plasticity model developed for pillar simulation is used to investigate the slip system activations predicted (*i.e.* for which the resolved shear stress is found to be greater than the CRSS, hence leading to non-zero slip rates). This methodology, of course, is more accurate than that in (ii) since it properly reflects the local multiaxial stress state.

The experimentally observed slip traces and the CP model predicted traces of slip system 2 ((010) [001]) are compared in Figure 7 (a)-(c) for pillar 1. The view angle, as sketched in the rotated coordinate system, is parallel to the Y-Z plane but rotated 52° about the X-axis, which is set as the same to the view angle in SEM observations. Good consistency between the CP model predicted and experimental observed slip traces (Figure 7 (b)) as quantified using the deviation angle 1°) justifies the conclusion of activation of slip system 2. This is further verified by the corresponding slip magnitude distribution (*i.e.* the slip contour (c) shows the direction of the potential slip plane trace) predicted using the CPFE model, and the total slip is found to be dominated by slip system 2. The single slip activation of Slip System 3 ((110) [001]) in pillar 2 is verified in Figure 7 (d)-(f). Therefore, the CPFE modelling successfully captures the single slip activations observed in pillars 1 and 2, together with the pillar nominal stress-strain responses, suggesting appropriate selection of material properties.

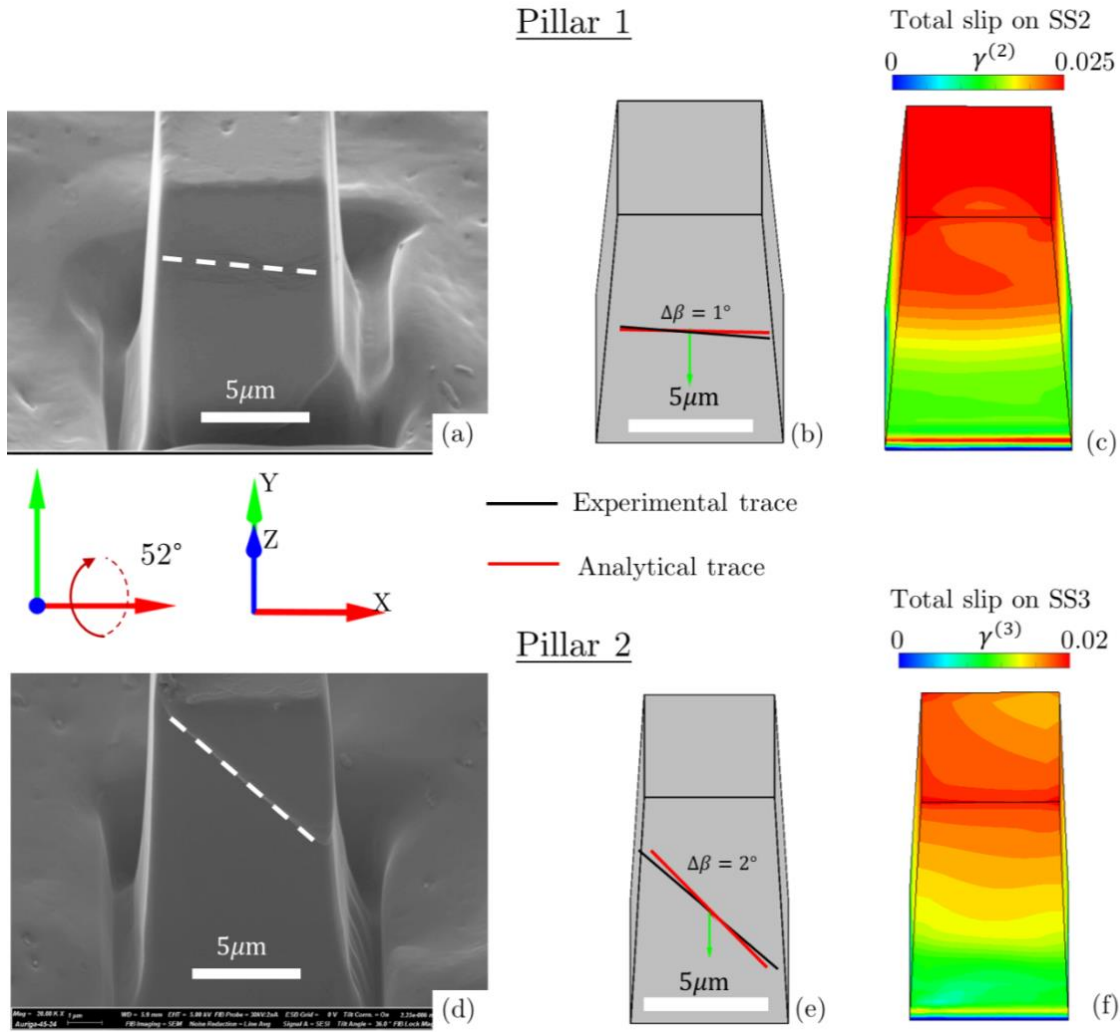

Figure 7. Slip traces (single) observed in the compression tests, and from CP prediction. (a)-(c) for pillar 1 (Slip System 2 solely activated) and (d)-(f) for pillar 2 (Slip System 3 solely activated). The view angles in the analytical solution and the CPFE simulation frame are chosen to be the same as in the SEM camera.

For pillar 3, the slip traces of slip system 8 ($(110)[\bar{1}11]/2$) and 10 ($(1\bar{1}0)[\bar{1}\bar{1}1]/2$) observed in the test are shown in Figure 8(a), and consistency between the experimental traces and the total slip from the CPFE model are demonstrated in Figure 8(b)-(c). For pillar 4, the slip traces of slip systems 2 $(010)[001]$ and 3 $(110)[001]$ and they are observed in Figure 8(d), and consistency between the experimental traces and the total slip from the CPFE model are demonstrated in Figure 8(e)-(f).

In both pillars, small deviations between the predicted and observed trace orientations are found, but the former are obtained based on the initial undeformed crystal orientation. Comparison of the experimental slip traces and the CPFE slip field plots show how double slip is required to enable shape change in the experimental pillars.

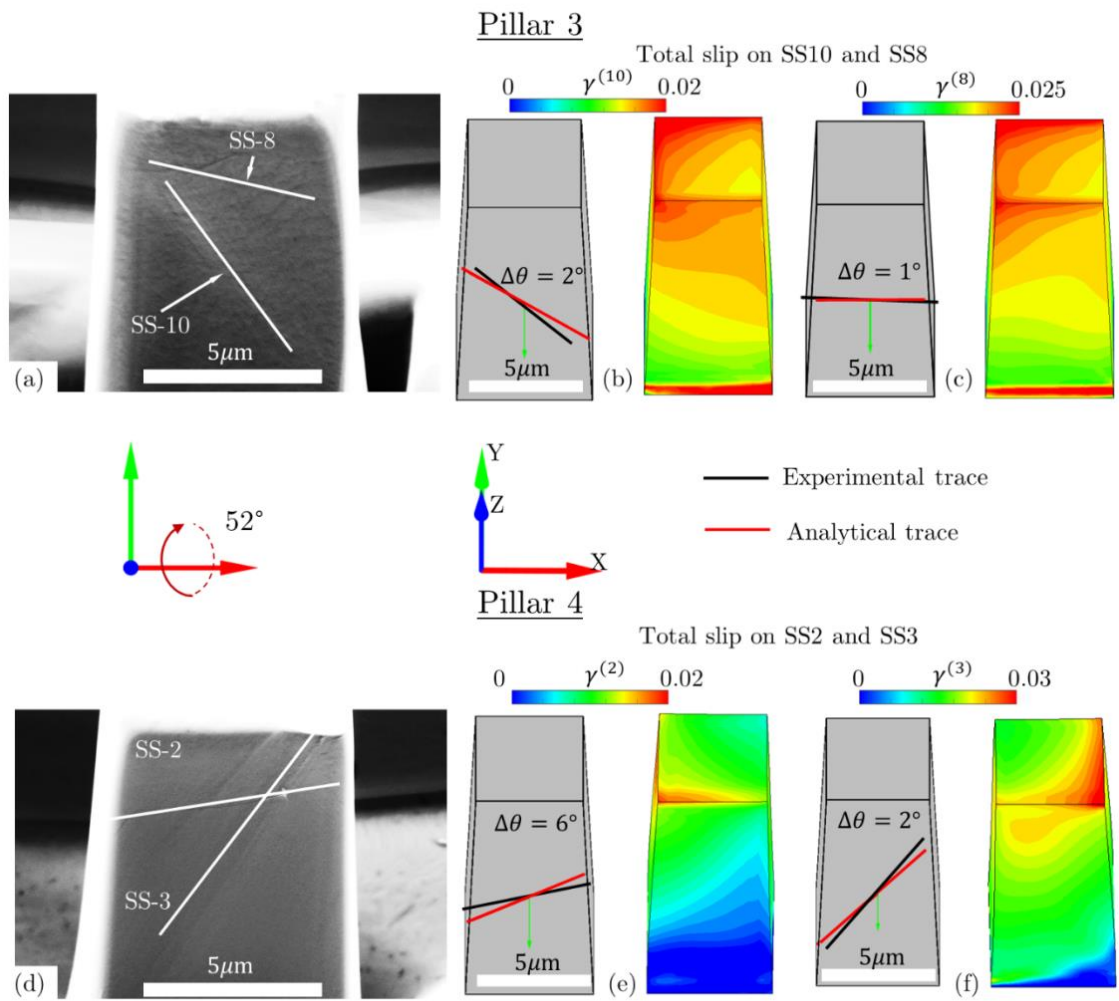

Figure 8. Double slip traces observed in the compression tests, and from CP modelling for (a)-(c) pillar 3 and (d)-(f) for pillar 4. The view angles in the analytical solution and the CPFE simulation frame are chosen the same as in the experimental observation (SEM camera). The slip distribution shows both slip system 8 and 10 are activated.

# 4  Establishment of the SAC305 slip properties

In the previous sections, the anisotropic elasto-plastic properties of the β-Sn single crystals have been established through a combined study of the micropillar test and CPFE simulation. Despite the predominating weight fraction (~96.5%) of Sn phase, the second phases (*i.e.* the intermetallic compounds $Ag_3Sn$ and $Cu_6Sn_5$) still significantly affect the thermomechanical behaviour (*e.g.* creep resistance [33]) of the SAC305 alloy. Therefore, it remains essential to establish the SAC305 alloy slip properties by incorporating its representative microstructure features, in addition to the knowledge of the β-Sn single crystals that has been established.

In this section, a multi-scale model is firstly presented to homogenize and incorporate the mechanical influences of the various phases present in SAC305. Once this is established, the rate sensitivity of a polycrystal representative volume element (RVE) SAC305 model is then examined at three typical temperatures, employing the single crystal β-Sn slip strengths and β-Sn slip properties obtained above. The results are then compared with those from independently reported experiments. The resulting homogenized model provides the link between single crystal properties and the macroscopic performance of SAC305 solders, when the knowledge of second phase IMC characteristic length scale is known.

## 4.1  Multi-scale homogenization of SAC305 alloy

The SAC305 model established by [33], where the β-Sn and IMC phases are explicitly represented, enabled the effects of IMC particle morphology, spacing and size in SAC305 to be investigated. However, such an approach with explicit model representation of the differing phases becomes largely intractable for representative solder analyses because of the multiple order of magnitude size differences between a typical solder joint and the IMC phase particles. Therefore, a physics-based, multi-length-scale homogenized model [7] is required to incorporate the effects of the key microstructure features of SAC305 alloys.

The detailed microstructure of SAC305 alloy and corresponding solder joints is complex and can be categorized at several scales, which are depicted in Figure 9. A SAC305 ball grid array (BGA) solder joint typically contains one to three β-Sn dendritic grains (shown Figure 9(a)). Within these, there are regions of β-Sn dendrite that are almost pure Sn with ~0.07 at% Ag and ~0.005 at% Cu [44] and regions of eutectic between dendrite arms that consist of β-Sn and micro- or nano-scale intermetallic compounds ($Ag_3Sn$ and $Cu_6Sn_5$) that originally formed in eutectic reactions during solder solidification, shown in (b)-(d). This leads to a hierarchy of length scale from (i) the β-Sn grains, to (ii) the dendrite arm spacing which determines the distance between eutectic regions, to (iii) the size and spacing of IMCs in the eutectic regions.

A multi-scale modelling approach is used to capture the mechanical performance of the characterised microstructures at the aforementioned scales. The β-Sn phase is assigned directly the material properties determined above from the β-Sn pillars, indicated in Figure 9 (h). The eutectic and eutectic + dendrite microstructures (shown in Figure 9 (g) and (f)) are incorporated by homogenization [25, 45, 46] utilising the methodology for SAC305 alloy proposed by [33], which eventually contributes to the establishment of an anisotropic SAC305 material model (Figure 9(e)). The latter work showed that provided the full microstructure (considering both grain morphology and crystallography) was explicitly modelled, the role of the IMC size could be incorporated through its influence on the slip strength and the hardening coefficients of β-Sn phase, thus providing a homogenized SAC305 polycrystal model. An overview of homogenization modelling of the eutectic regions and the β-Sn dendrites can be found in Figure A 4 which is fully detailed in [33]. This homogenization has been used in the present paper, and its validity examined in a later section by comparing the predicted strain-rate sensitive response of a polycrystal representative volume element (RVE) and independent experimental tests at various temperatures. The established anisotropic SAC305 material properties which depend upon both β-Sn crystallographic orientation and IMC size are used to investigate the dislocation-mediated damage of solder joints under thermal cycling, and the homogenization methodology validated by comparisons of the plastic strain predicted by the homogenized CPFE model with an independent continuum model [47] .

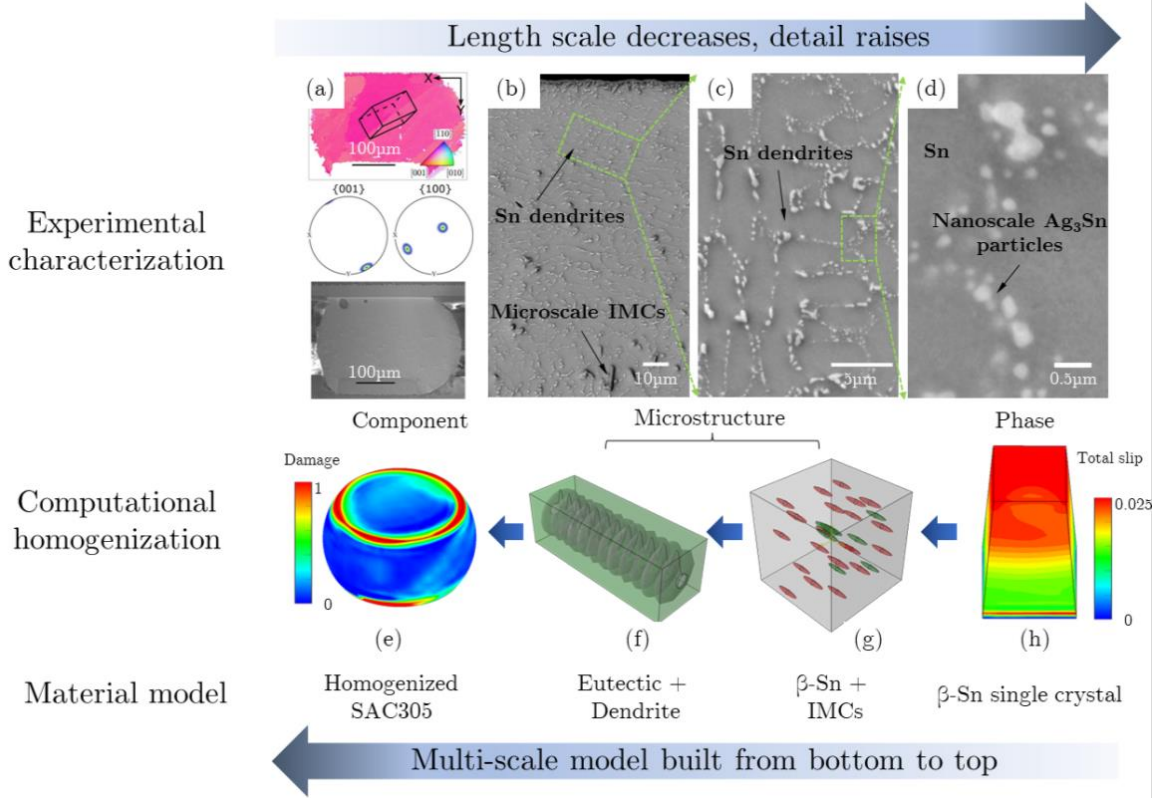

Figure 9. Road map of establishing homogenized SAC305 single crystal model by using the multi-scale modelling approach. (a)-(d) the experimental observation of the microstructures in a SAC305 solder. (f)-(i) the corresponding multi-scale modelling approaches that builds up the homogenized SAC305 grain detained in [33] .

The initial intrinsic strength of the <001> family slip system $\tau_{c0}^{<001>}$ within a given homogenized SAC305 single crystal in the homogenised model depends upon the IMC particle size, where the functional relationship has been quantified for 25°C in [33] to be

$$\tau_{c0}^{<001>} = -0.8 \times \bar{r}_{IMC} + 5.4 \quad (\text{MPa}) \tag{1}$$

where $\bar{r}_{IMC}$ is the average characteristic IMC size (in μm) within the SAC305 crystal, considering both $Ag_3Sn$ and $Cu_6Sn_5$ phases. The slip strengths of the other slip systems are obtained by applying the CRSS ratios provided in [39]. The IMC phases also contribute to hardening and was incorporated in the homogenised model through the statistically stored dislocation (SSD) density ($\rho_{ssd}$) hardening evolution. The latter is determined from the effective plastic strain rate $\dot{p}$ and current accumulated plastic strain $p$ [48], where the recovery term reflects that shear resistance stabilizes when a large deformation develops [49, 50]. The SSD density accumulation rate at an instant is given by:

$$\dot{\rho}_{ssd} = \lambda_1 \dot{p} - \lambda_2 p \quad (\mu m^{-2} s^{-1}) \qquad (2)$$

Here, $\lambda_1$ and $\lambda_2$ control the slip system hardening and recovery rates respectively, and are found to be related to the average IMC radius (in µm)[33] by:

$$\lambda_1 = -260 \times \bar{r}_{IMC} + 1215 \quad (\mu m^{-2})$$
$$\lambda_2 = -4.7 \times \bar{r}_{IMC} + 15.2 \quad (\mu m^{-2} s^{-1}) \qquad (3)$$

The IMC size within a given solder alloy is physically determined by the undercooling reached prior to β-Sn nucleation during the solidification process [14, 51, 52], and coarsening during isothermal aging and thermal cycling in service [2, 53, 54]. The representative average IMC radius (equivalent circular radius, ECR) for SAC305 BGA joints [53, 54] are estimated as

$$0.11 \, \mu m < \bar{r}_{IMC} < 1.88 \, \mu m \qquad (4)$$

where the minimum and maximum values were quantified as soldered and at a sufficiently long duration, respectively.

The temperature *sensitivities* of both the elastic moduli and slip strengths (the latter deduced from the temperature *sensitivity* of yield stresses) are taken to be those determined by [5] but of course used to be consistent with the single crystal slip properties determined in this paper, which are directly obtained from the Sn single crystal pillar tests and presented above, for room temperature 25°C. The additional properties needed for the homogenised Sn-IMC 'crystals' ($\lambda_1$ and $\lambda_2$) are given in Table 4, along with the temperature sensitivities of the elastic constants (given in Table 3 at 25°C) and the homogenised slip system strengths (for which Eqn.(1) gives the dependence on IMC size). All other properties required remain unchanged and are given in Table 3.

Table 4. Temperature-dependent material properties for the homogenized SAC305 crystal (temperature T in °C)

| $E_1$ (GPa) | $E_3$ (GPa) | $\tau_{c0}^{<001>}$ (MPa) |
|---|---|---|
| $25.3 - 0.083 \times (T - 273)$ | $68.3 - 0.072 \times (T - 273)$ | $4.9 - 0.02 \times (T - 273)$ |

## 4.2 Polycrystal simulation and comparison to independent experiments

A polycrystal representative volume element (RVE) is developed which is cubic with edge length 300 μm, as shown in Figure 10(a), and comprises 125 grains with random crystallographic orientations (which reflects a representative weak texture commonly observed in the polycrystal samples [55]), and each individual grain is assigned with the material properties of SAC305 single crystal with hardening coefficients which include dependency on IMC size and distribution. Displacement-controlled uniaxial tension is imposed on the SAC305 polycrystal RVE, and the stress-strain response is then evaluated at various strain rates and temperatures. The boundary conditions for the isothermal uniaxial tension are schematically shown in Figure 10(b). The left, bottom and back faces of the RVE are constrained along the X, Y and Z directions respectively, and a displacement is uniformly imposed upon the top face along the Y-direction, $U_y$=U. The macroscopic strain is calculated as ε=U/L, and four loading rates $\dot{\epsilon}$ =$10^{-2}$ s$^{-1}$, $10^{-3}$ s$^{-1}$, $10^{-4}$ s$^{-1}$ and $10^{-5}$ s$^{-1}$ are realized. The macroscopic stress is calculated as the ratio of the sum of reaction forces along the top face and the area *i.e.*, σ=∑F$_y$/A. The temperature, and IMC size dependent stress-strain response of the polycrystal RVE is first assessed for two extreme values of the average IMC sizes ($\bar{r}_{IMC}$=0.11 μm and $\bar{r}_{IMC}$=1.88 μm) assigned to the homogenized crystal plasticity model.

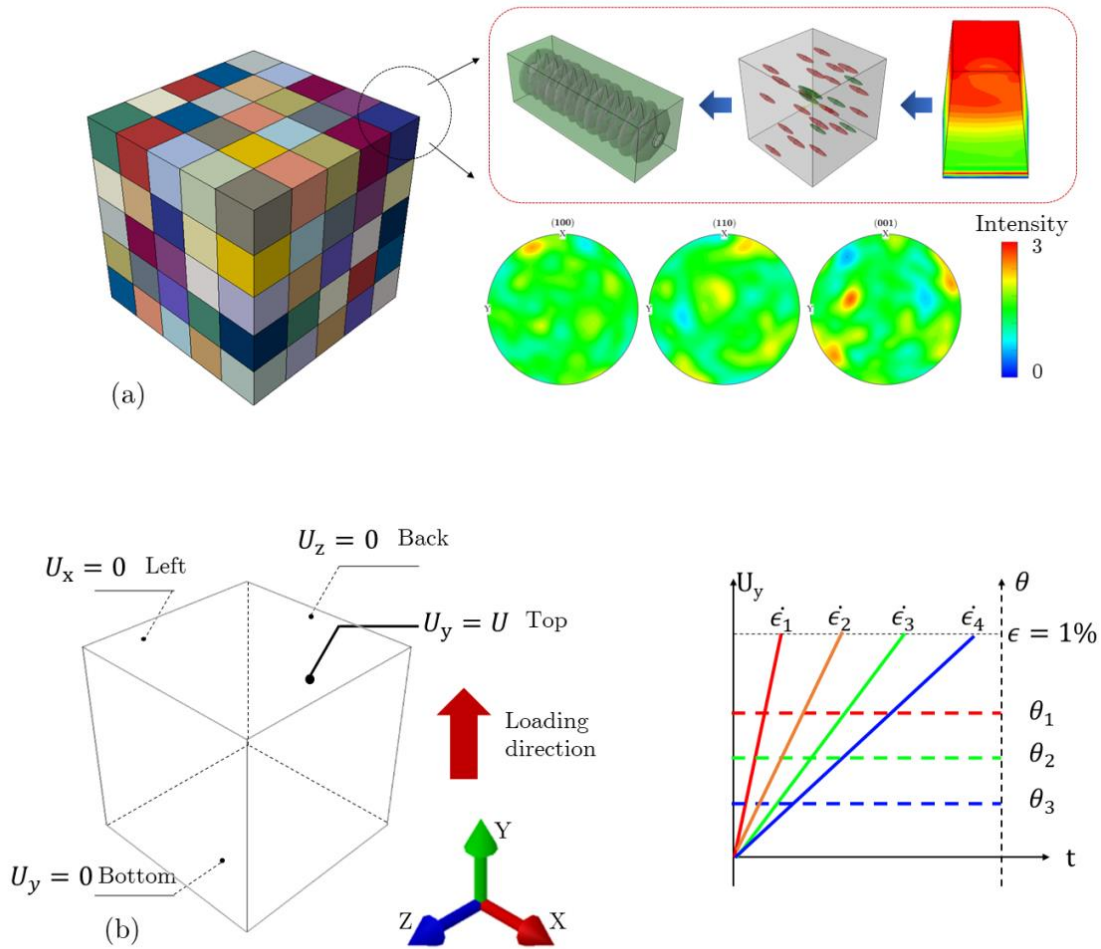

Figure 10. (a) Polycrystal representative volume element (RVE) for of the SAC305 alloy with individual grains homogenized from the multi-scale approach (b) the isothermal, displacement-controlled uniaxial boundary conditions imposed to the RVE with three temperature and four loading rates applied, respectively.

Three temperatures which cover the range for typical solder thermal cycling [56, 57] of 25℃, 75℃ and 125℃ are considered in the polycrystal RVE modelling. The resulting predicted temperature-dependent average flow stresses versus applied strain rate for the polycrystal model are shown in Figure 11 (for the four applied strain rates). In addition, a set of independent tensile test results [5, 58-62] have also been included showing generally good agreement with the predicted results. While the IMC size range relevant to the tests was not reported, it is likely to be consistent with the range selected ($\bar{r}_{IMC}$ =0.11 μm and $\bar{r}_{IMC}$ =1.88 μm). The upper and lower extremes shown for each temperature correspond to these two IMC sizes which are relevant to the temperature conditions considered [54]. The predicted curves are obtained by utilising knowledge of the IMC size range (see Eqn. (2)) for each temperature of interest and using the corresponding homogenised SAC305 slip strength and hardening described in Eqns. (2) and (3) by applying the two representative IMC sizes. The discrepancy in grain size and morphology [63], texture [64], sample size [65] and aging time [66] could give rise to the scattering of flow stresses, which in turn highlights the dependence of the thermomechanical behaviour of SAC305 alloy upon its microstructure.

Strong temperature sensitivity is predicted at all strain rates investigated here. In addition, the predicted flow stresses are reasonably consistent with independent tension tests on polycrystal SAC305 samples [5], whose data are indicated as the circular symbols in the figures. In addition, rate sensitivity is also predicted for polycrystal SAC305 samples, which arises from time-dependent dislocation jump and escape events [35] along a slip system. Both the sensitivity to loading rate and temperature are important in the temperature-dependent creep behaviour of Sn-based solder joints [7, 8, 67], and are considered to be significant mechanisms for driving solder joint damage under in-service conditions [68]. The predicted flow stresses of the homogenized polycrystal RVE are consistent with the independent experimental measurement, thus justifying the property determination and the multi-scale homogenised SAC305 model.

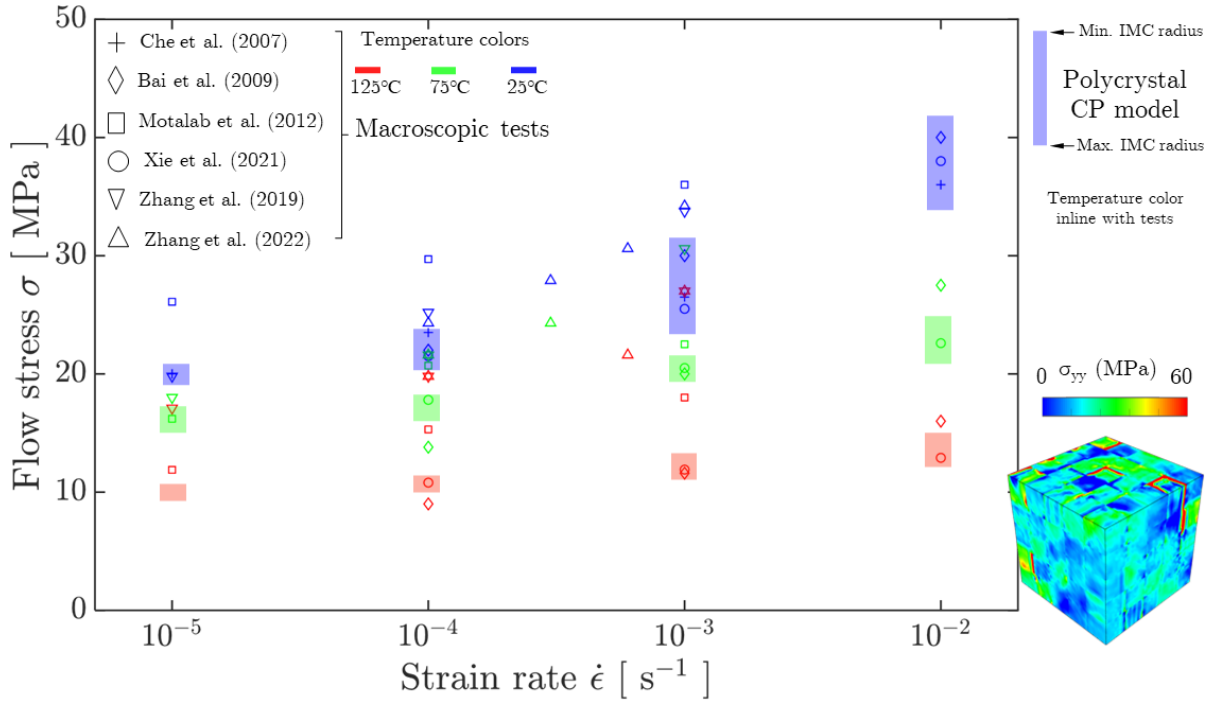

Figure 11. Predicted temperature-dependent flow stress of the SAC305 polycrystals subject to strain rates $\dot{\epsilon} = 10^{-2}$ s$^{-1}$, $10^{-3}$ s$^{-1}$, $10^{-4}$ s$^{-1}$ and $10^{-5}$ s$^{-1}$, respectively, and temperatures 25℃, 75℃ and 125℃ using the homogenised CPFE model. The upper and lower limits reflect the IMC size range (*i.e.* the average radius ranging 0.11-1.88 μm) observed at the temperatures indicated. Independent macroscopic SAC305 tensile test data (symbols) included for comparison. An example polycrystal YY-stress distribution is shown inset for the strain rate $10^{-2}$ s$^{-1}$ at 25℃ at strain of 1%.

## 5  Conclusion

This paper presents a step-by-step experimental calibration and model development of rate-sensitive mechanical performance of β-Sn containing alloys, and multi-scale homogenisation which considers the role of anisotropy and key microstructural features (*e.g.* IMC phases and their distribution in eutectic regions).

Major conclusions include:

i) The differing β-Sn slip system strengths and rate-sensitivity governing parameters have been quantified through comparison of micro-mechanical testing experiments with crystal plasticity simulations.

ii) Together these enable the construction of a multi-scale homogenised model that enables prediction of polycrystalline multi-phase SAC305 alloys over a useful temperature range (25 to 125°C) to predict the performance of components and support the understanding of the failure mechanisms of solder joints during thermal cycling.

iii) A mechanistic link from fundamental β-Sn single-crystal properties through to polycrystal multi-phase SAC305 alloy behaviour over temperature and strain rate has been demonstrated, providing a useful predictive capability for SAC305 solder joints under thermal cycling conditions for a given IMC size, or the evolving IMC size if known.

## Acknowledgement

All authors acknowledge the financial support by the Engineering and Physical Sciences Research Council for funding through the grant EP/R018863/1. FPED wishes to acknowledge gratefully the provision of Royal Academy of Engineering/Rolls-Royce research chair funding. TBB wishes to acknowledge the Royal Academy of Engineering for funding his research fellowship.

[69] A. Zamiri, T.R. Bieler, F. Pourboghrat, Anisotropic Crystal Plasticity Finite Element Modeling of the Effect of Crystal Orientation and Solder Joint Geometry on Deformation after Temperature Change, Journal of Electronic Materials 38(2) (2009) 231-240. http://doi.org/10.1007/s11664-008-0595-0.

# Appendix A

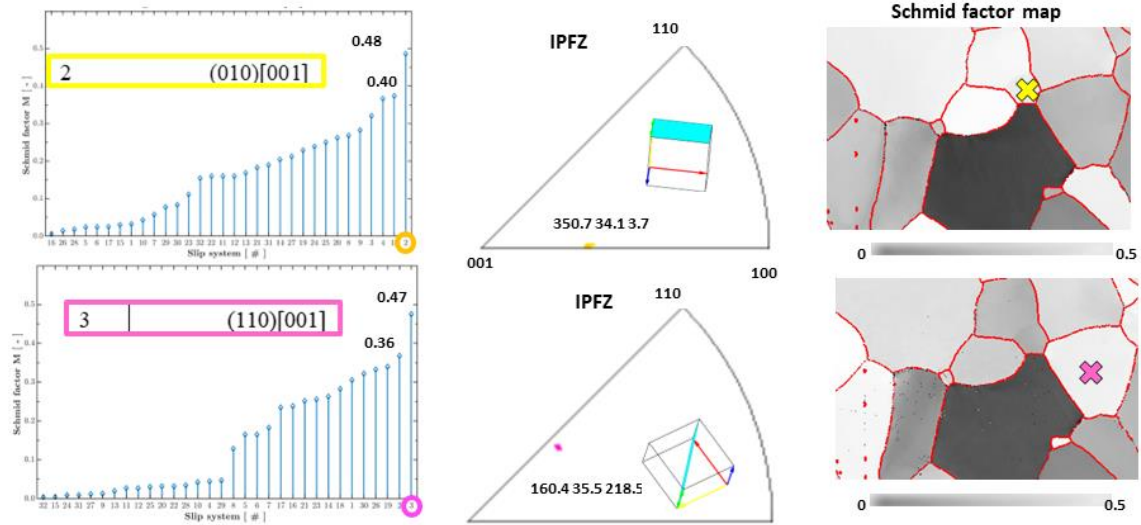

Figure A 1. Schmid factor analysis of the two pillars. The activation sequence of the slip systems is based on prior works by Fujiwara et al.[41] and Zhou et al. [39]. The common Sn slip systems with CRSS ratios ($\tau^{CRSS}/\tau^{[001]}$) are included in the table. The CRSS ratios are estimated by Zamiri et al. [69] that consider different crystal orientations for single-crystal samples under uniaxial tensile test from Zhou et al. [39].

Table A 1. The common Sn slip systems [41] with CRSS (the critical resolved shear stress) ratios given. The values of CRSS ratio are estimated from by Zamiri et al. [69]

| Family No. | | Slip system | CRSS ratio |
|---|---|---|---|
| 1 | 1 | (100)[001] | 1 |
| | 2 | (010)[001] | 1 |
| 2 | 3 | (110)[001] | 1 |
| | 4 | ($1\bar{1}0$)[001] | 1 |
| 3 | 5 | (100)[010] | 1.05 |
| | 6 | (010)[100] | 1.05 |
| 4 | 7 | (110)[1$\bar{1}$1]/2 | 1.1 |
| | 8 | (110)[$\bar{1}$11]/2 | 1.1 |
| | 9 | (1$\bar{1}$0)[111]/2 | 1.1 |
| | 10 | (1$\bar{1}$0)[$\bar{1}$$\bar{1}$1]/2 | 1.1 |
| 5 | 11 | (110)[1$\bar{1}$0] | 1.2 |
| | 12 | (1$\bar{1}$0)[110] | 1.2 |
| 6 | 13 | (010)[101] | 1.25 |
| | 14 | (010)[10$\bar{1}$] | 1.25 |
| | 15 | (100)[011] | 1.25 |
| | 16 | (100)[01$\bar{1}$] | 1.25 |
| 7 | 17 | (001)[100] | 1.3 |
| | 18 | (001)[010] | 1.3 |
| 8 | 19 | (001)[110] | 1.4 |
| | 20 | (001)[1$\bar{1}$0] | 1.4 |
| 9 | 21 | (101)[10$\bar{1}$] | 1.5 |
| | 22 | (10$\bar{1}$)[101] | 1.5 |
| | 23 | (011)[01$\bar{1}$] | 1.5 |
| | 24 | (01$\bar{1}$)[011] | 1.5 |
| 10 | 25 | (121)[$\bar{1}$01] | 1.5 |
| | 26 | (1$\bar{2}$1)[$\bar{1}$01] | 1.5 |
| | 27 | ($\bar{1}$21)[101] | 1.5 |
| | 28 | ($\bar{1}$$\bar{2}$1)[101] | 1.5 |
| | 29 | (211)[0$\bar{1}$1] | 1.5 |
| | 30 | ($\bar{2}$11)[0$\bar{1}$1] | 1.5 |

| | | |
|---|---|---|
| 31 | $(\bar{2}11)[011]$ | 1.5 |
| 32 | $(2\bar{1}1)[011]$ | 1.5 |

# Appendix B

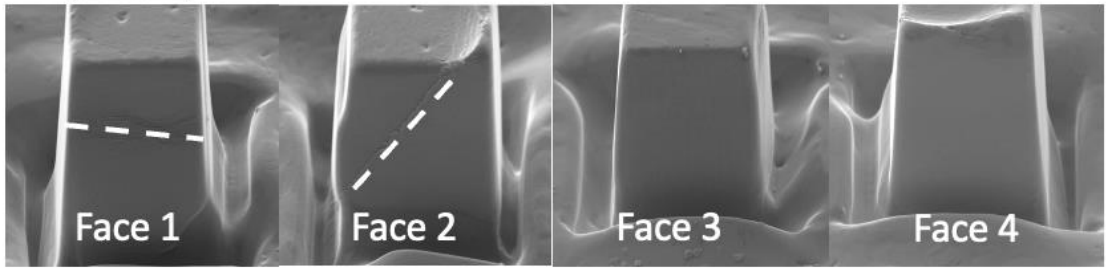

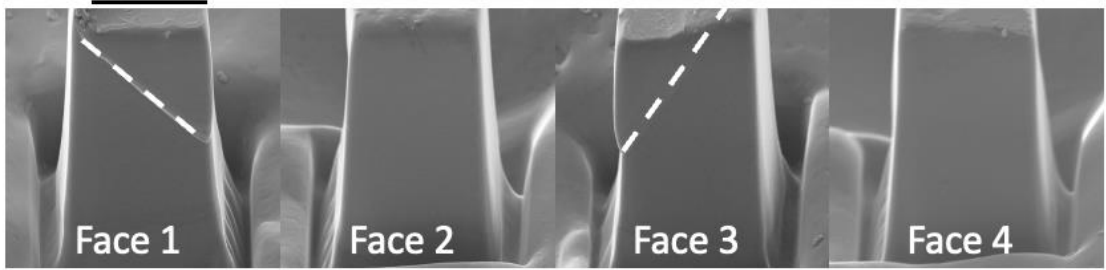

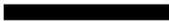

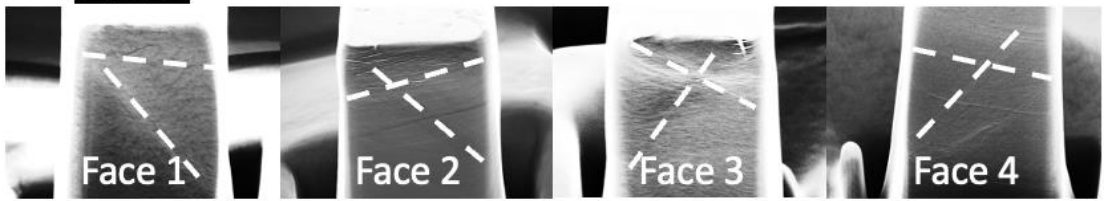

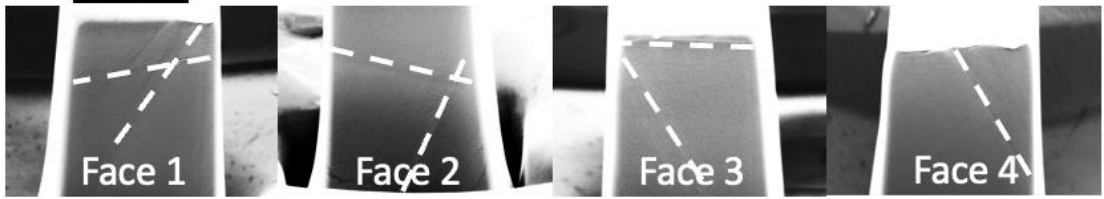

Figure A 2. SEM micrographs of deformed micropillars for pillar 1, pillar 2, pillar 3 and pillar 4 showing all four surfaces with experimental slip traces annotated. The micrographs for pillar 3 and pillar 4 are modified from ref [27].

# Appendix C

The plastic strain rate $\dot{\boldsymbol{\varepsilon}}^p$ is given by the summation of shear rates $\dot{\gamma}^{(i)}$ for all active slip systems as

$$\dot{\boldsymbol{\varepsilon}}^p = sym(\sum_{i=1}^{M} \dot{\gamma}^{(i)} \boldsymbol{s}^{(i)} \otimes \boldsymbol{n}^{(i)}) \quad (5)$$

where $M$ is the total number of active slip systems (up to 32 for the body centred tetragonal (BCT) crystal as β-Sn), and $\boldsymbol{s}^{(i)}$ and $\boldsymbol{n}^{(i)}$ are the slip direction and normal direction vectors of the $i$th slip system, respectively. The slip rate $\dot{\gamma}^{(i)}$ along the $i^{th}$ slip system is governed by thermally activated dislocation escape from obstacles, which gives:

$$\dot{\gamma}^{(i)} = \rho_{ssdm} b^2 v_D \exp\left(-\frac{\Delta H}{k\theta}\right) \sinh\left(\frac{\Delta V}{k\theta}\left|\tau^{(i)} - \tau_c^{(i)}\right|\right) \quad (6)$$

where $\rho_{ssdm}$ is the initial mobile statistically stored dislocation density, $b$ the magnitude of the Burgers vector, $v_D$ the dislocation jump frequency, $\Delta H$ the activation energy with associated activation volume $\Delta V$, $k$ the Boltzmann constant, and $\tau^{(i)}$ and $\tau_c^{(i)}$ are the resolved shear stress and critical resolved shear stress on $i$th slip system, respectively. The critical resolved shear stress $\tau_c^{(i)}$ evolves with dislocation density, which is correlated to the effective plastic strain, according to a hardening law.

A length scale dependant, dislocation-based hardening law is adopted for the evolution of shear resistance along the $i$th slip system:

$$\tau_c^{(i)} = \tau_{c0} + Gb\sqrt{\rho_{SSD} + \rho_{GND}} \quad (7)$$

where $\tau_{c0}$ is the intrinsic critical resolved shear stress, $G$ the shear modulus, and $\rho_{SSD}$ and $\rho_{GND}$ are the density of statistically stored (SSD) and geometry necessary dislocations (GND), respectively. The evolution of SSD density is related to the effective plastic strain rate $\dot{p}$ and accumulated plastic strain $p$ in a recovery rule, where recovery corresponds to the fact that shear resistance saturates when a large deformation applies:

$$\dot{\rho}_{SSD} = \lambda_1 \dot{p} - \lambda_2 p \tag{8}$$

Here, $\lambda_1$ and $\lambda_2$ control the slip system hardening and recovery rates respectively and are determined from micro-pillar tests as detailed below. The effective plastic strain $p$ is given by:

$$p = \sqrt{\frac{2}{3}\boldsymbol{\varepsilon}^p : \boldsymbol{\varepsilon}^p} \tag{9}$$

where $\boldsymbol{\varepsilon}^p$ is the plastic strain tensor.

The GND density term $\rho_{GND}$ in eq.(7) is established from the plastic strain gradients that accommodate lattice curvature. The components of the GND density tensor are calculated from Nye's dislocation tensor $\boldsymbol{\Lambda}$ that originates from the curl of plastic deformation gradient $\boldsymbol{F}^p$:

$$\boldsymbol{\Lambda} = \nabla \times \boldsymbol{F}^p = \sum_{i=1}^{32} \rho_{Gs}^i \boldsymbol{b}^i \otimes \boldsymbol{s}^i + \rho_{Get}^i \boldsymbol{b}^i \otimes \boldsymbol{t}^i + \rho_{Gen}^i \boldsymbol{b}^i \otimes \boldsymbol{n}^i \tag{10}$$

A scalar GND density is then given by the by L$_2$-norm minimization of the edge and screw dislocation density components of Nye's tensor:

$$\rho_{GND} = \sqrt{\sum_{i=1}^{32} (\rho_{Gs}^i)^2 + (\rho_{Get}^i)^2 + (\rho_{Gen}^i)^2} \tag{11}$$

## Appendix D

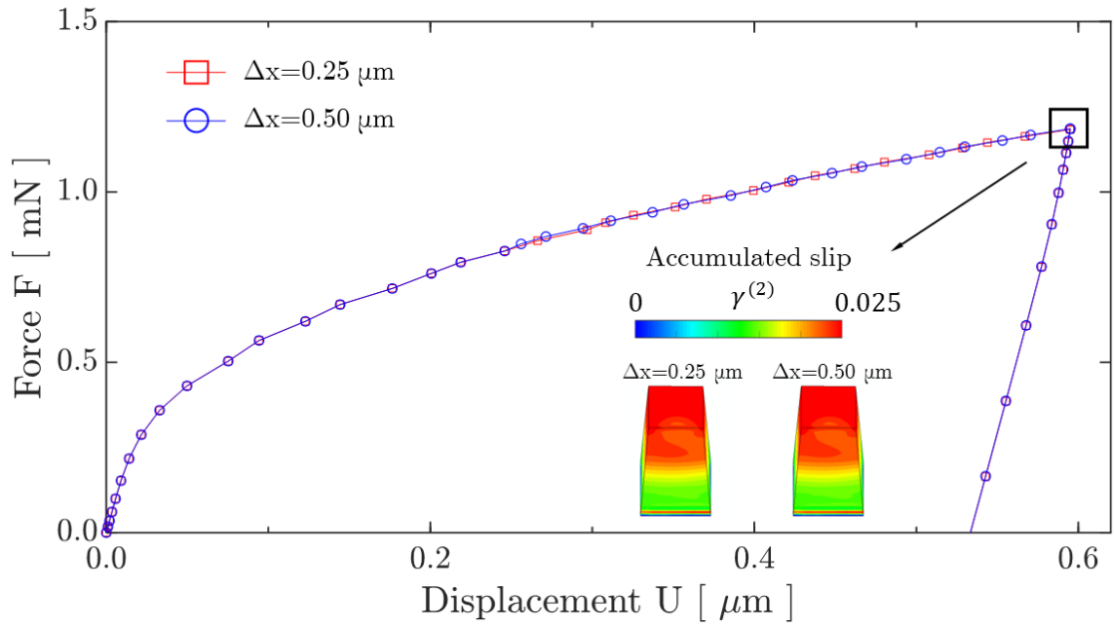

Figure A 3. The mesh size sensitivity study of the CPFE model. Pillar 3 is taken as an example.

## Appendix E

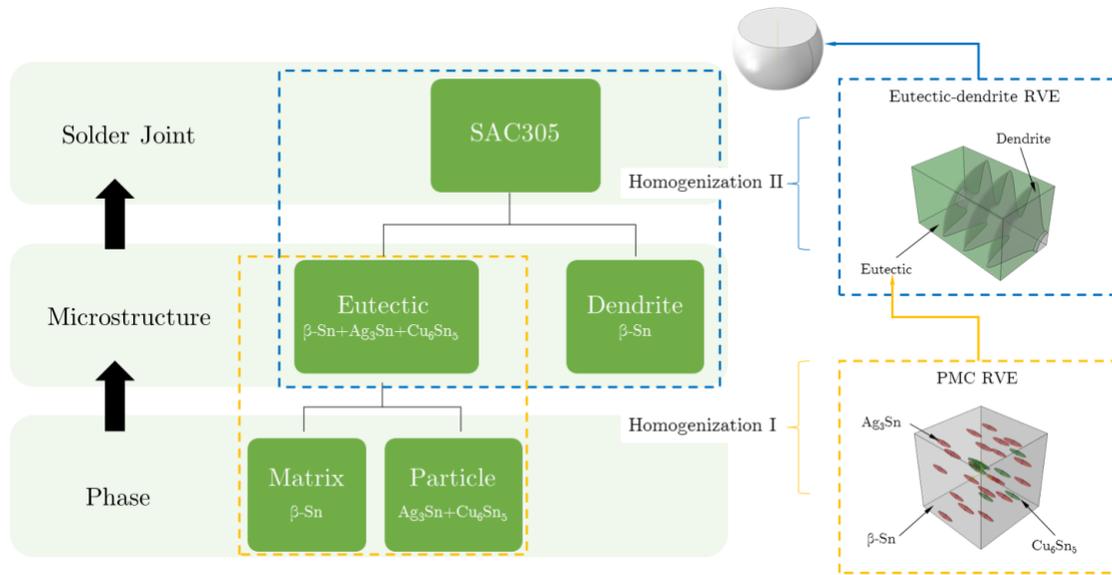

Figure A 4. the multi-scale modelling strategy for establishing the material properties of SAC305 alloys used in solder joints [33].